\documentclass[%
 aip,
 amsmath,amssymb,
 reprint,%
 groupedaddress 
]{revtex4-1}

\usepackage{graphicx}
\usepackage{float}
\usepackage{dcolumn}
\usepackage{bm}

\usepackage[caption=false]{subfig}
\usepackage{siunitx, soul, xcolor}
\usepackage{adjustbox}

\begin{document}
\preprint{APS/123-QED}

\title{Atomic Layer Molecular Beam Epitaxy of Kagome Magnet RMn$_6$Sn$_6$ (R = Er, Tb) Thin Films}

\author{Shuyu Cheng}
\affiliation{Department of Physics, The Ohio State University, Columbus, Ohio 43210, United States}
\author{Igor Lyalin}
\affiliation{Department of Physics, The Ohio State University, Columbus, Ohio 43210, United States}
\author{Wenyi Zhou}
\affiliation{Department of Physics, The Ohio State University, Columbus, Ohio 43210, United States}
\author{Roland K. Kawakami}
\email{kawakami.15@osu.edu}
\affiliation{Department of Physics, The Ohio State University, Columbus, Ohio 43210, United States}

\begin{abstract}
Kagome lattices have garnered substantial interest because their band structure consists of topological flat bands and Dirac cones. 
The RMn$_6$Sn$_6$ (R = rare earth) compounds are particularly interesting because of the existence of large intrinsic anomalous Hall effect (AHE) which originates from the gapped Dirac cones near the Fermi level.
This makes RMn$_6$Sn$_6$ an outstanding candidate for realizing the high-temperature quantum anomalous Hall effect.
The growth of RMn$_6$Sn$_6$ thin films is beneficial for both fundamental research and potential applications.
However, most of the studies on RMn$_6$Sn$_6$ have focused on bulk crystals so far, and the synthesis of RMn$_6$Sn$_6$ thin films has not been reported so far.
Here we report the atomic layer molecular beam epitaxy growth, structural and magnetic characterizations, and transport properties of ErMn$_6$Sn$_6$ and TbMn$_6$Sn$_6$ thin films.
It is especially noteworthy that TbMn$_6$Sn$_6$ thin films have out-of-plane magnetic anisotropy, which is important for realizing the quantum anomalous Hall effect.
Our work paves the avenue toward the control of the AHE using devices patterned from RMn$_6$Sn$_6$ thin films.
\end{abstract}

\flushbottom
\maketitle
\thispagestyle{empty}

Materials with layered kagome structures provide an ideal platform for studying the interplay between band structure and magnetism. 
In momentum space, a two-dimensional kagome lattice manifests Dirac cones at the K points~\cite{kuroda2017evidence, ye2018massive}, saddle points at the M points~\cite{kang2022twofold, teng2023magnetism}, and a flat band across the Brillouin zone~\cite{kang2020topological, kang2020dirac}.
The introduction of magnetism and spin-orbital coupling further opens a gap at the band touching points, leading to topologically nontrivial electronic states~\cite{sun2011nearly, guo2009topological, bolens2019topological}.
In real space, non-trivial spin textures (e.g. magnetic skyrmions) have been observed in kagome magnets~\cite{hou2017observation, li2023discovery}.

Recently, RMn$_6$Sn$_6$ (R = rare earth) compounds have emerged as a new family of topological kagome magnets~\cite{yin2020quantum, ma2021rare}.
The Mn atoms in RMn$_6$Sn$_6$ form Mn$_3$ kagome layers and the spin ordering of the Mn$_3$ layers is largely affected by the R atoms.
Previous studies on bulk crystals report that R = Gd, Tb, Dy, and Ho give rise to ferromagnetic coupling of Mn spins, while R = Er, Tm, and Lu lead to antiferromagnetic coupling of Mn spins at zero field~\cite{venturini1991magnetic, malaman1999magnetic, clatterbuck1999magnetic}.
Rare earth elements in RMn$_6$Sn$_6$ compounds also provide one additional knob for tuning the magnetocrystalline anisotropy (MCA) which is connected to the Dirac gap opening~\cite{cheng2023visualization}.
Large Berry-curvature-induced intrinsic anomalous Hall effect (AHE) may originate from the gapped Dirac cones close to the Fermi level, because the intrinsic anomalous Hall conductivity (AHC) is connected to the Dirac gap opening $\Delta$ and the Dirac cone position $E_D$ through~\cite{sinitsyn2007anomalous}:
\begin{equation}
\sigma^{int}=\frac{e^2}{h}\frac{\Delta}{2E_D} \label{eq1}
\end{equation}
Notably, TbMn$_6$Sn$_6$ bulk crystals exhibit out-of-plane magnetic anisotropy, which leads to a 34\,meV Dirac gap opening at 130\,meV above the Fermi level~\cite{yin2020quantum}, and band structure calculations predict the quantum anomalous Hall effect (QAHE) if the Fermi level could be tuned into the gap.
Experimentally, the observation of a giant intrinsic AHC of $\sigma^{int}$=0.14\,e$^2$/$h$ per kagome layer in TbMn$_6$Sn$_6$ bulk crystals~\cite{yin2020quantum} makes the RMn$_6$Sn$_6$ family of materials attractive for QAHE.
To this end, the growth of RMn$_6$Sn$_6$ thin films provides opportunities for tuning the band structure and Fermi level through epitaxial strain, chemical doping, or voltage gating.
However, all of the reported works have focused on bulk crystals of RMn$_6$Sn$_6$, while studies on their thin film counterparts have been missing.
Therefore, the development of epitaxial RMn$_6$Sn$_6$ thin films is highly desired.

In this paper, we report the growth of (0001)-oriented thin films of ErMn$_6$Sn$_6$ and TbMn$_6$Sn$_6$ using atomic layer molecular beam epitaxy (AL-MBE) and characterize their magnetic and transport properties. The films are grown on Pt(111) buffer layers on Al$_2$O$_3$(0001) substrates.
Various methods are employed to characterize the structure of our samples, including \textit{in-situ} reflection high energy electron diffraction (RHEED), X-ray diffraction (XRD), and atomic force microscopy (AFM).
We demonstrate that the magnetic properties of RMn$_6$Sn$_6$ thin films can be tuned through the selection of rare earth element $R$.
ErMn$_6$Sn$_6$ thin films favor in-plane (IP) magnetization, while TbMn$_6$Sn$_6$ thin films undergo an out-of-plane (OOP) to IP spin-reorientation transition (SRT) as the temperature increases above 300\,K.
Lastly, we find that the longitudinal resistance of RMn$_6$Sn$_6$/Pt bilayers do not follow a simple parallel resistor model but instead depends on the electron mean free path and can be explained qualitatively by the Fuchs-Sondheimer model.

\begin{figure}[h]
    \subfloat[\label{fig:Structure}]{
        \includegraphics[width=0.4\textwidth]{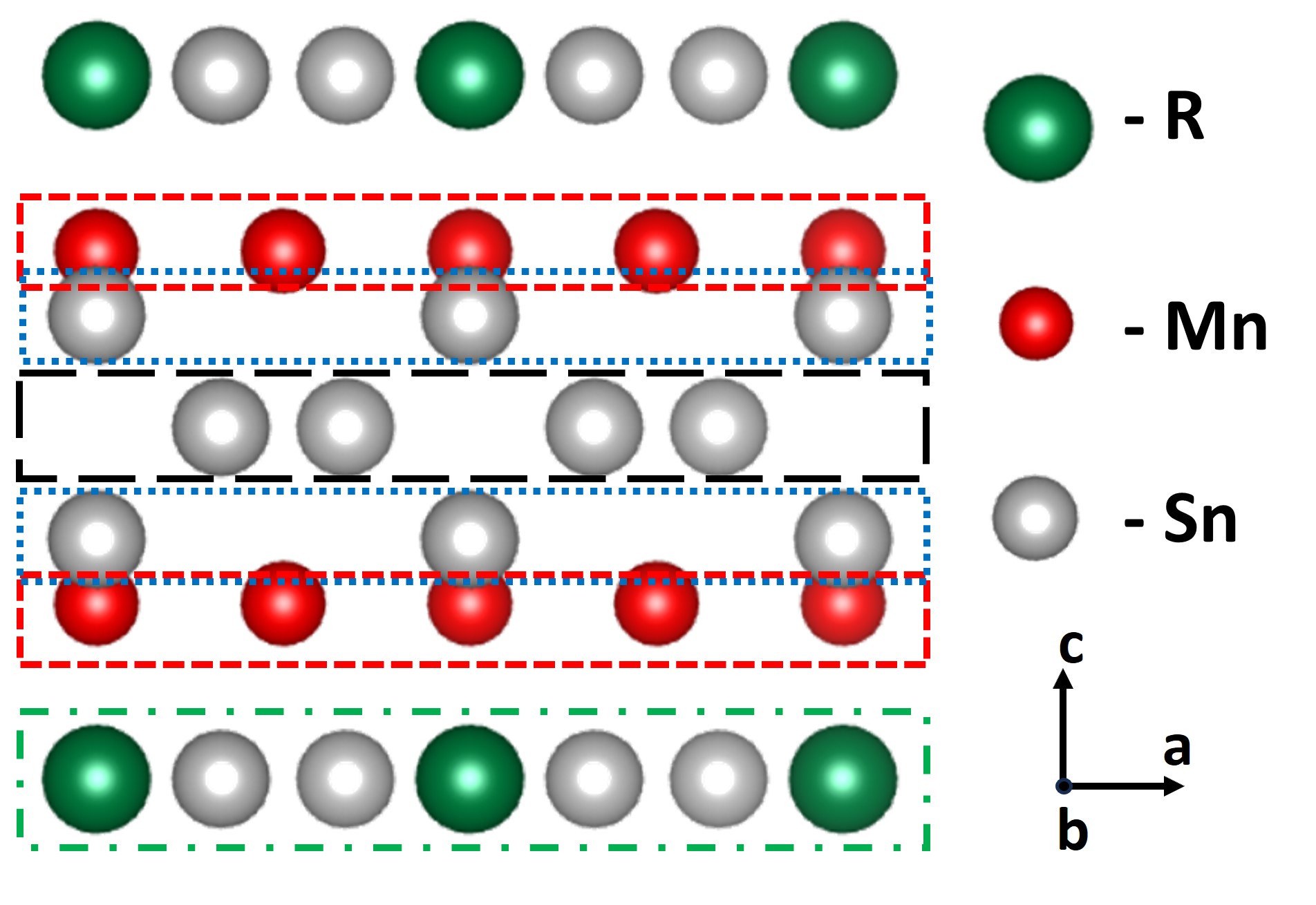}
        }\hfill
    \subfloat[\label{fig:Structure}]{
        \includegraphics[width=0.45\textwidth]{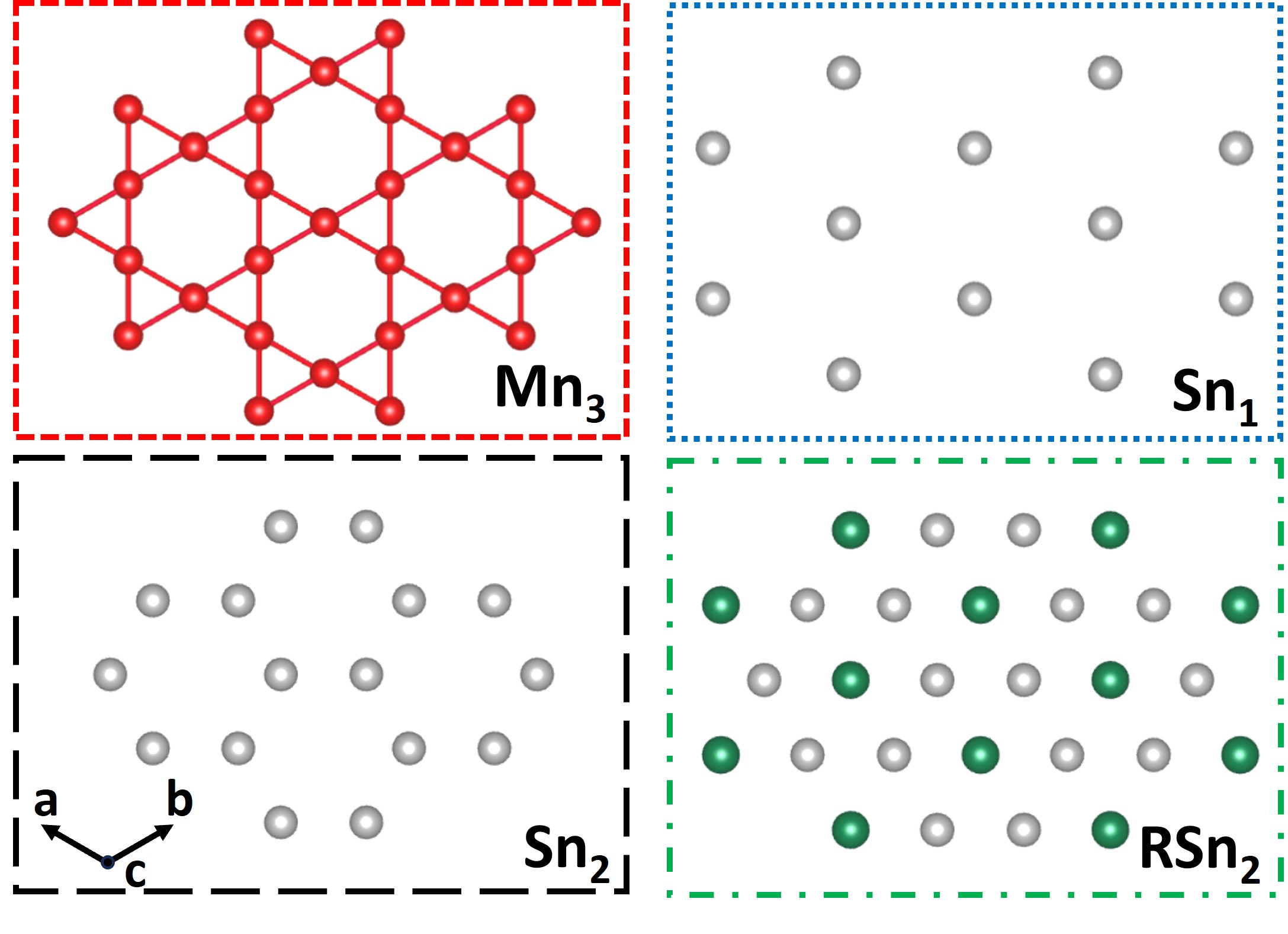}
        }\hfill
    \subfloat[\label{fig:RHEED_Er166}]{
        \includegraphics[width=0.23\textwidth]{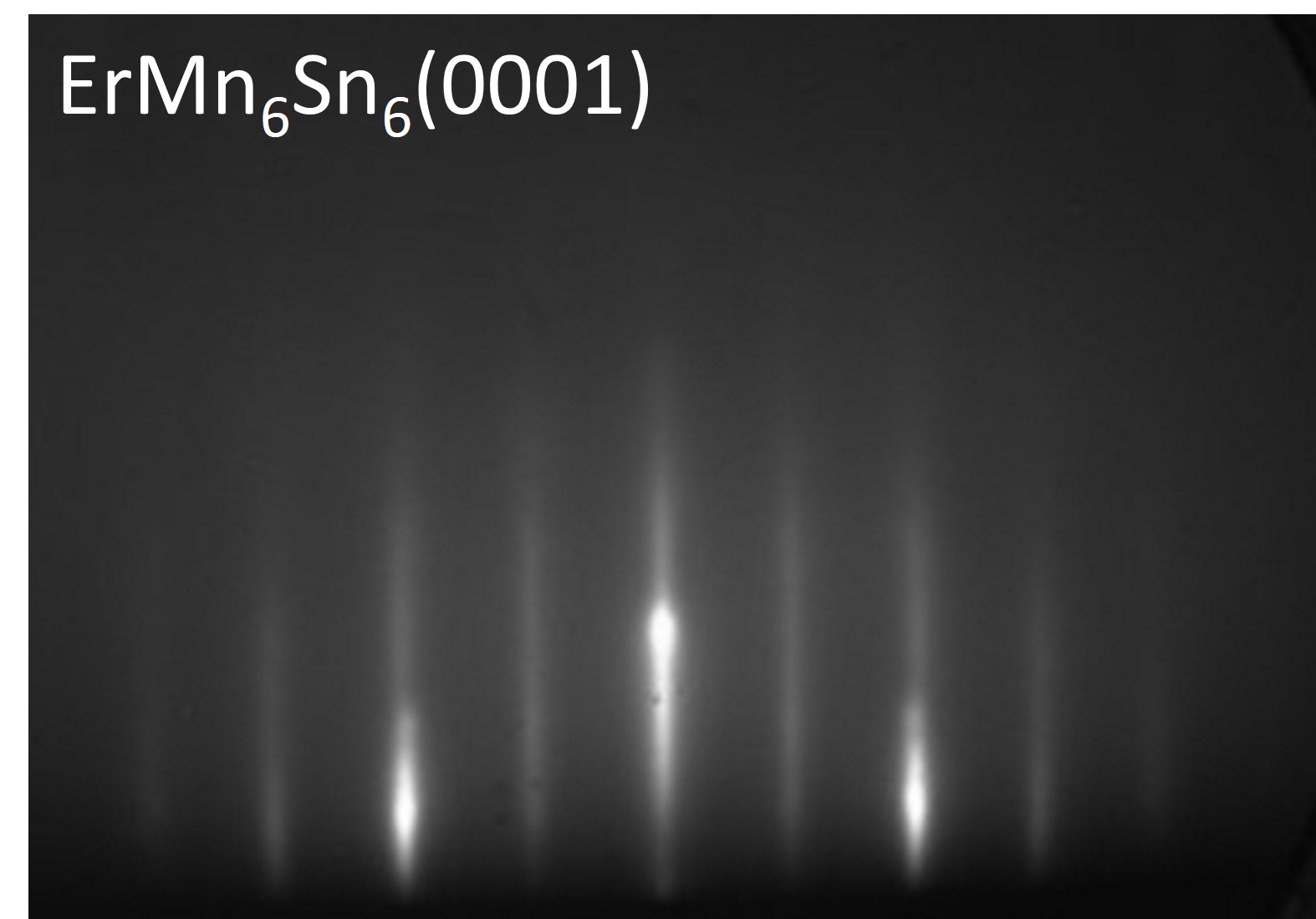}
        }
    \subfloat[\label{fig:RHEED_Tb166}]{
        \includegraphics[width=0.23\textwidth]{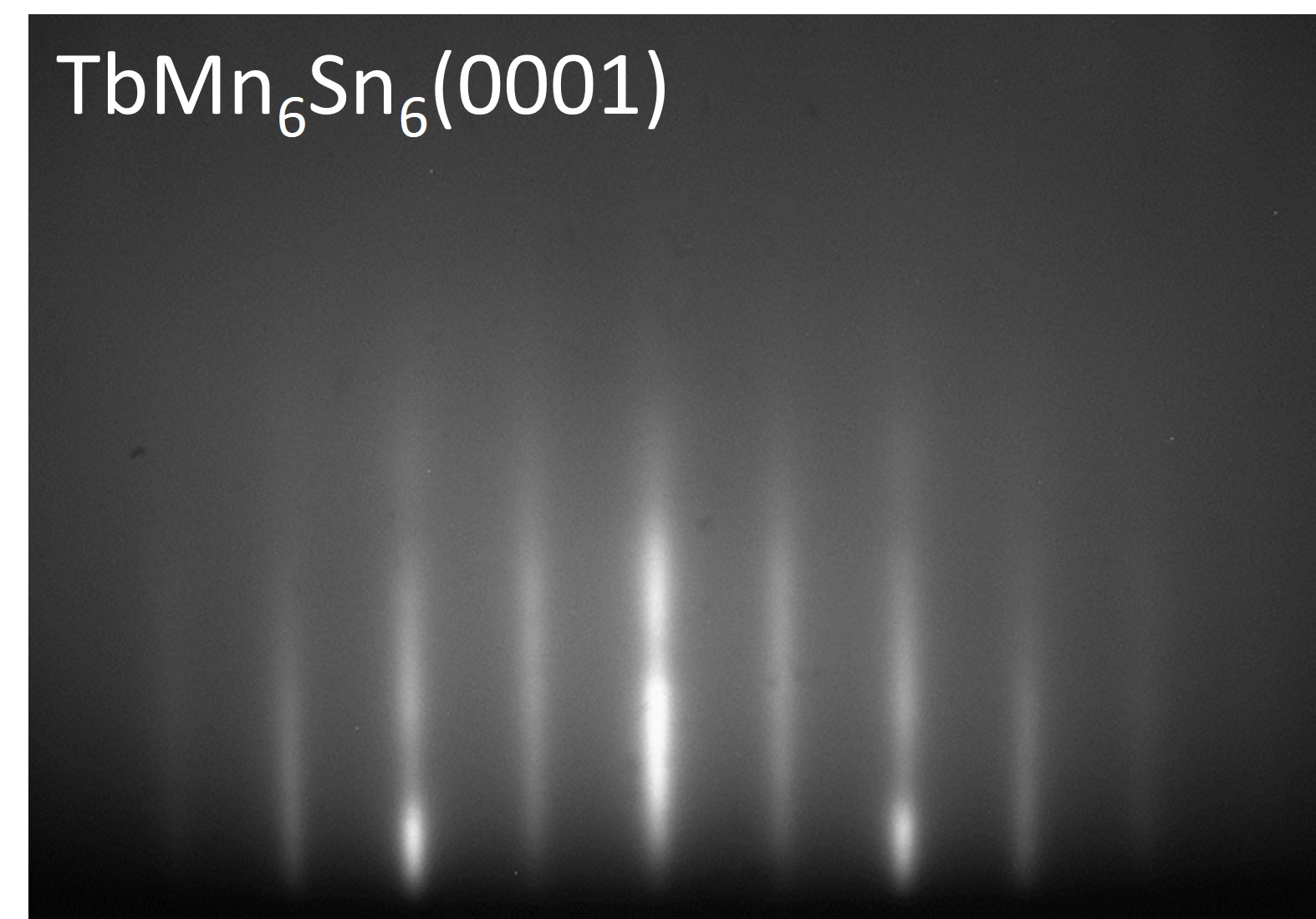}
        }
    \caption{\label{fig:Growth} AL-MBE Growth of RMn$_6$Sn$_6$ thin films.
    (a). Lattice structure of RMn$_6$Sn$_6$ viewed from the side. 
    In each unit cell, RMn$_6$Sn$_6$ consists of alternating stacking of RSn$_2$, Mn$_3$, Sn$_1$, Sn$_2$, Sn$_1$, Mn$_3$ layers (from bottom to top) along c-axis.
    (b). In-plane lattice structure of kagome Mn$_3$ (top left), triangular Sn$_1$ (top right), honeycomb Sn$_2$ (bottom left), and RSn$_2$ (bottom right) layers.
    (c, d). \textit{in-situ} RHEED pattern of 20\,nm ErMn$_6$Sn$_6$ sample grown on Pt(111) buffer layer on Al$_2$O$_3$(0001) substrate.
    (d). \textit{in-situ} RHEED pattern of 20\,nm TbMn$_6$Sn$_6$ sample grown on Pt(111) buffer layer on Al$_2$O$_3$(0001) substrate.
    } 
\end{figure}

The crystal structure of RMn$_6$Sn$_6$ is displayed in Fig.~\ref{fig:Growth}, featuring a hexagonal lattice (space group $P6/mmm$, lattice constants a = 5.5\,\AA~and c = 9.0\,\AA) with atomic layers stacked along the c-axis~\cite{malaman1988nouveaux}.
In each unit cell, there are two kagome Mn$_3$ layers closely neighboured by two Sn$_1$ layers, with one honeycomb Sn$_2$ layer inserted between two Sn$_1$ layers, and one RSn$_2$ layer inserted between Mn$_3$ kagome layers.
Inspired by this stacking sequence, we develop an AL-MBE recipe to grow ErMn$_6$Sn$_6$ and TbMn$_6$Sn$_6$ thin films on epitaxial Pt(111) buffer layers on Al$_2$O$_3$(0001) substrates. 
In AL-MBE, atomic layers are grown sequentially by opening and closing the appropriate MBE source shutters with precise timing.
A similar growth strategy allowed us to synthesize thin films of another kagome magnet, Fe$_3$Sn$_2$, which also has a hexagonal layered lattice structure~\cite{cheng2022atomic}.

The base pressure of the growth chamber is 1$\times$10$^{-9}$\,Torr.
The Al$_2$O$_3$(0001) substrates (MTI corporation) are annealed in air at 1000\,$^{\circ}$C for 3 hours, then degassed \textit{in-situ} at 500\,$^{\circ}$C for 30 minutes to prepare a flat and clean surface.
A 5\,nm epitaxial Pt(111) buffer layer is subsequently grown on Al$_2$O$_3$(0001) following the recipe described in our previous work~\cite{cheng2022atomic}.
The ErMn$_6$Sn$_6$(0001) and TbMn$_6$Sn$_6$(0001) thin films are grown on Pt(111) buffer layers at 100\,$^{\circ}$C and 80\,$^{\circ}$C, respectively using the AL-MBE recipe described below.
First, the kagome Mn$_3$ layer and hexagonal Sn$_1$ layer are co-deposited with flux ratio Mn:Sn = 3:1, then the honeycomb RSn$_2$ is deposited with flux ratio Er:Sn (or Tb:Sn) = 2:1, then 
another hexagonal Sn$_1$ layer and kagome Mn$_3$ layer are co-deposited, then one Sn$_2$ layer is deposited.
The aforementioned shutter growth sequence is repeated until the desired thickness (typically 20\,nm) of RMn$_6$Sn$_6$ is achieved.
Typical deposition rates are 1.6\,\AA/min, 1.6\,\AA/min, 3.9\,\AA/min, and 2.8\,\AA/min for Er, Tb, Mn, and Sn, respectively.
After growth, a 5\,nm CaF$_2$ layer is deposited to prevent sample oxidation.

The \textit{in-situ} RHEED patterns of ErMn$_6$Sn$_6$ and TbMn$_6$Sn$_6$ are monitored during the growths.
The RHEED patterns of a 20\,nm ErMn$_6$Sn$_6$ thin film and a 20\,nm TbMn$_6$Sn$_6$ thin film along Al$_2$O$_3$[11$\bar{2}$0] direction are shown in Fig.~\ref{fig:RHEED_Er166} and~\ref{fig:RHEED_Tb166}, respectively.
The RHEED patterns are generally streaky, suggesting the morphology of our RMn$_6$Sn$_6$ thin films is dominated by two-dimensional terraces with finite sizes, although some three-dimensional features also exist as spots.
The RHEED pattern suggests the in-plane lattice constants for our samples are $a_{ErMn_{6}Sn_{6}}$= 5.47\,\AA, and $a_{TbMn_{6}Sn_{6}}$= 5.56\,\AA, respectively, and the in-plane epitaxial relationship is RMn$_6$Sn$_6$[1$\bar{1}$00]$\parallel$Pt[110]$\parallel$Al$_2$O$_3$[11$\bar{2}$0].

\begin{figure*}[ht]
    \subfloat[\label{fig:XRD_Er166}]{
        \includegraphics[width=0.58\textwidth]{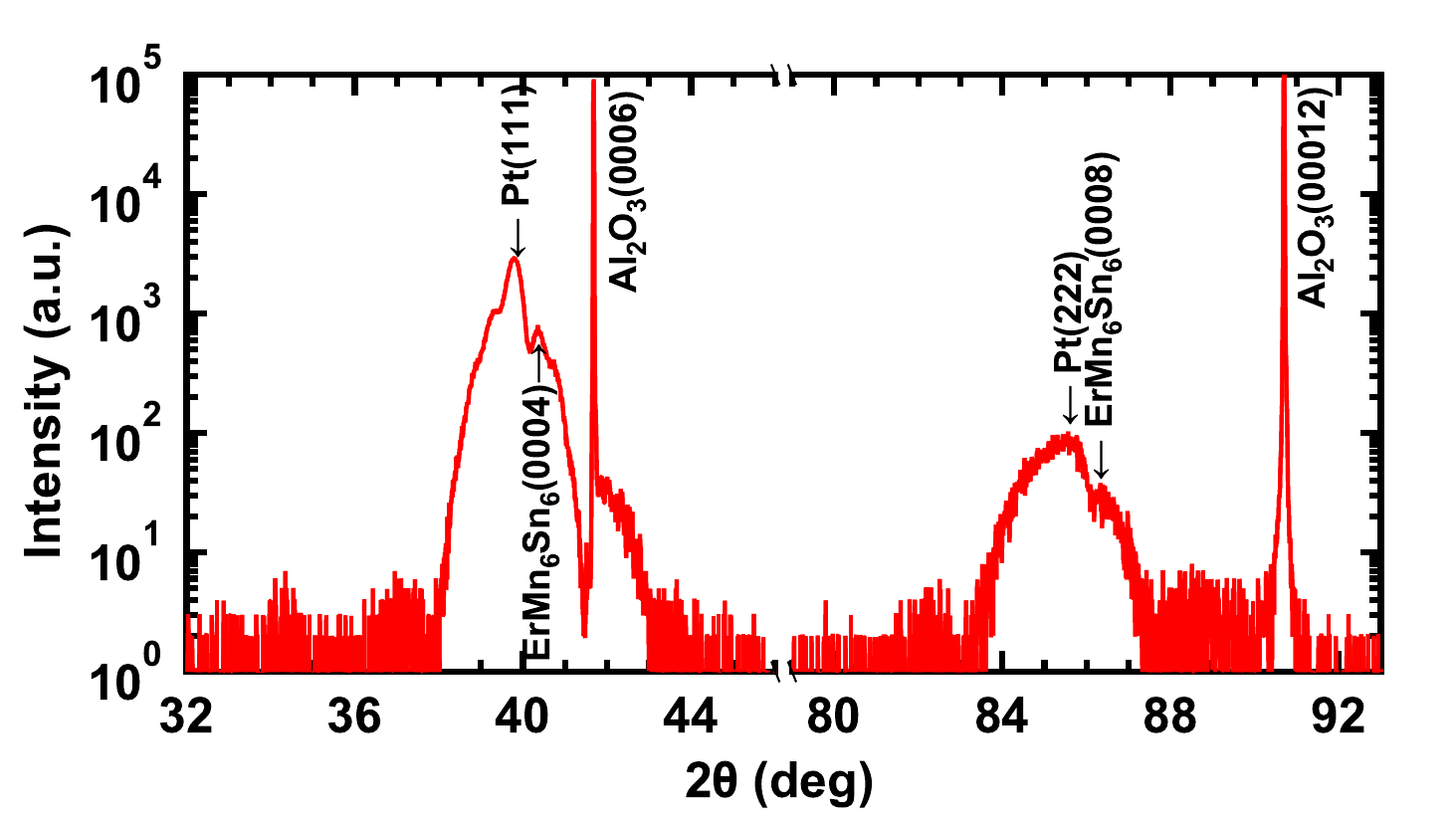}
        }
    \subfloat[\label{fig:AFM_Er166}]{
        \includegraphics[width=0.42\textwidth]{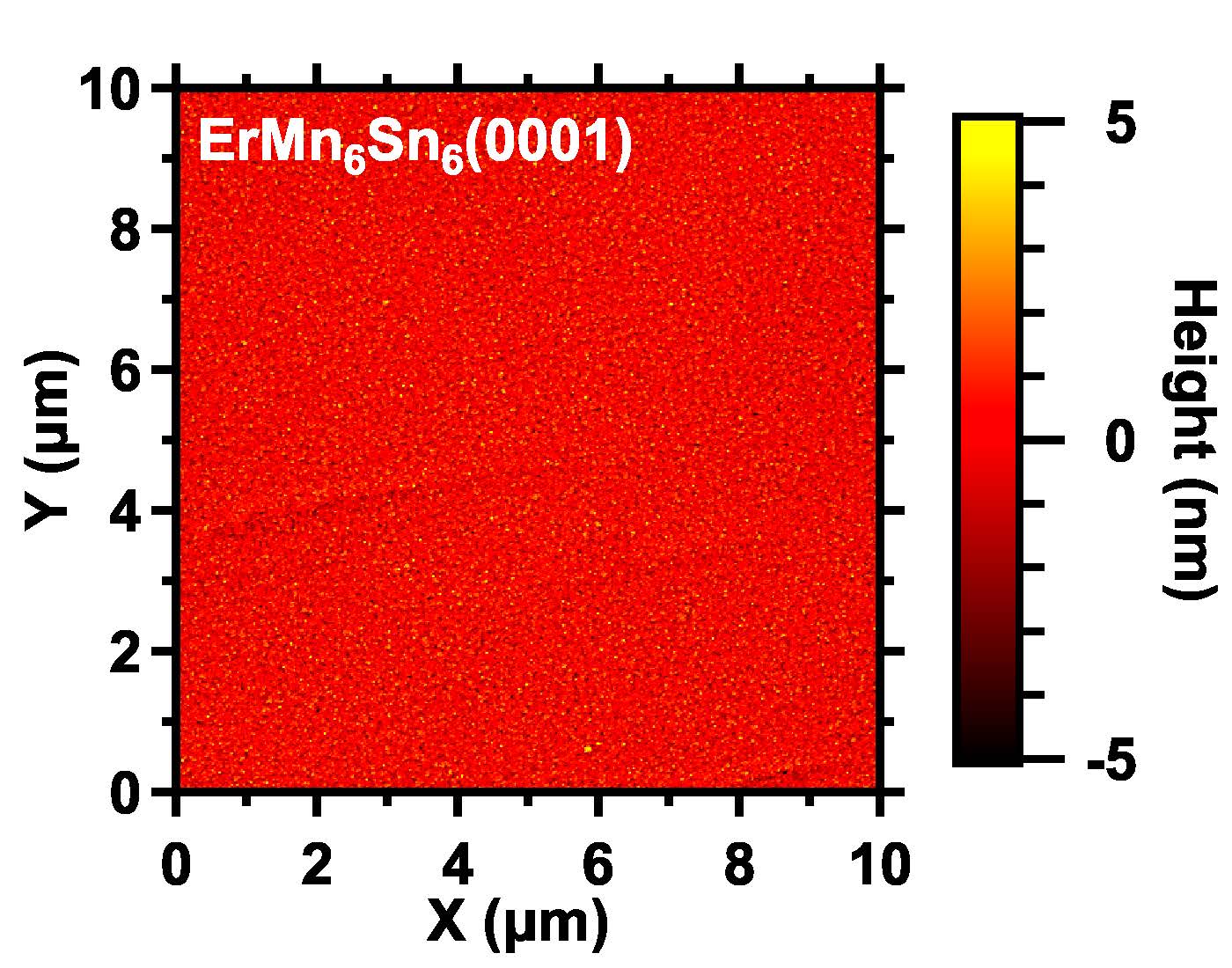}
        }\hfill
    \subfloat[\label{fig:XRD_Tb166}]{
        \includegraphics[width=0.58\textwidth]{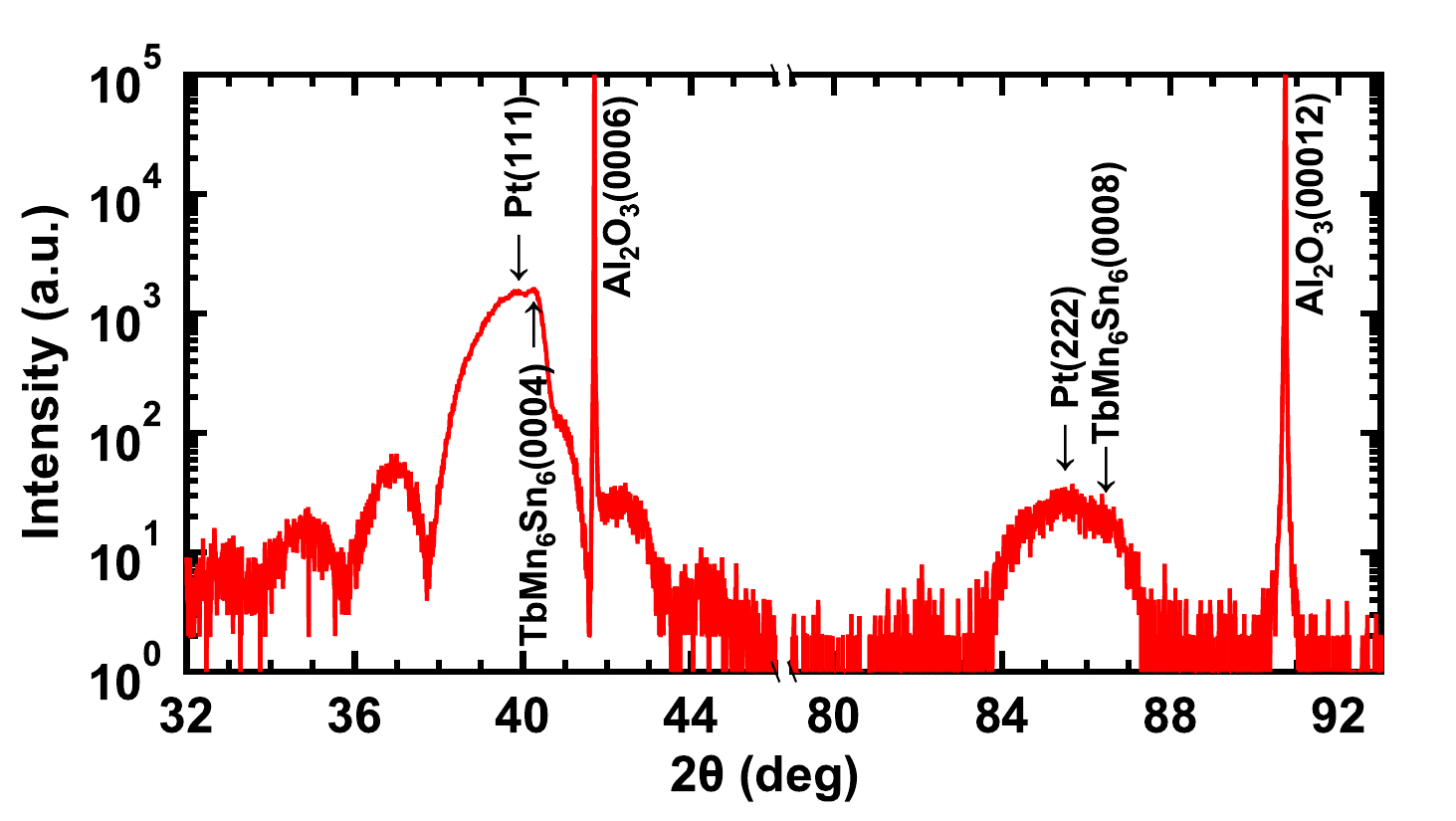}
        }
    \subfloat[\label{fig:AFM_Tb166}]{
        \includegraphics[width=0.42\textwidth]{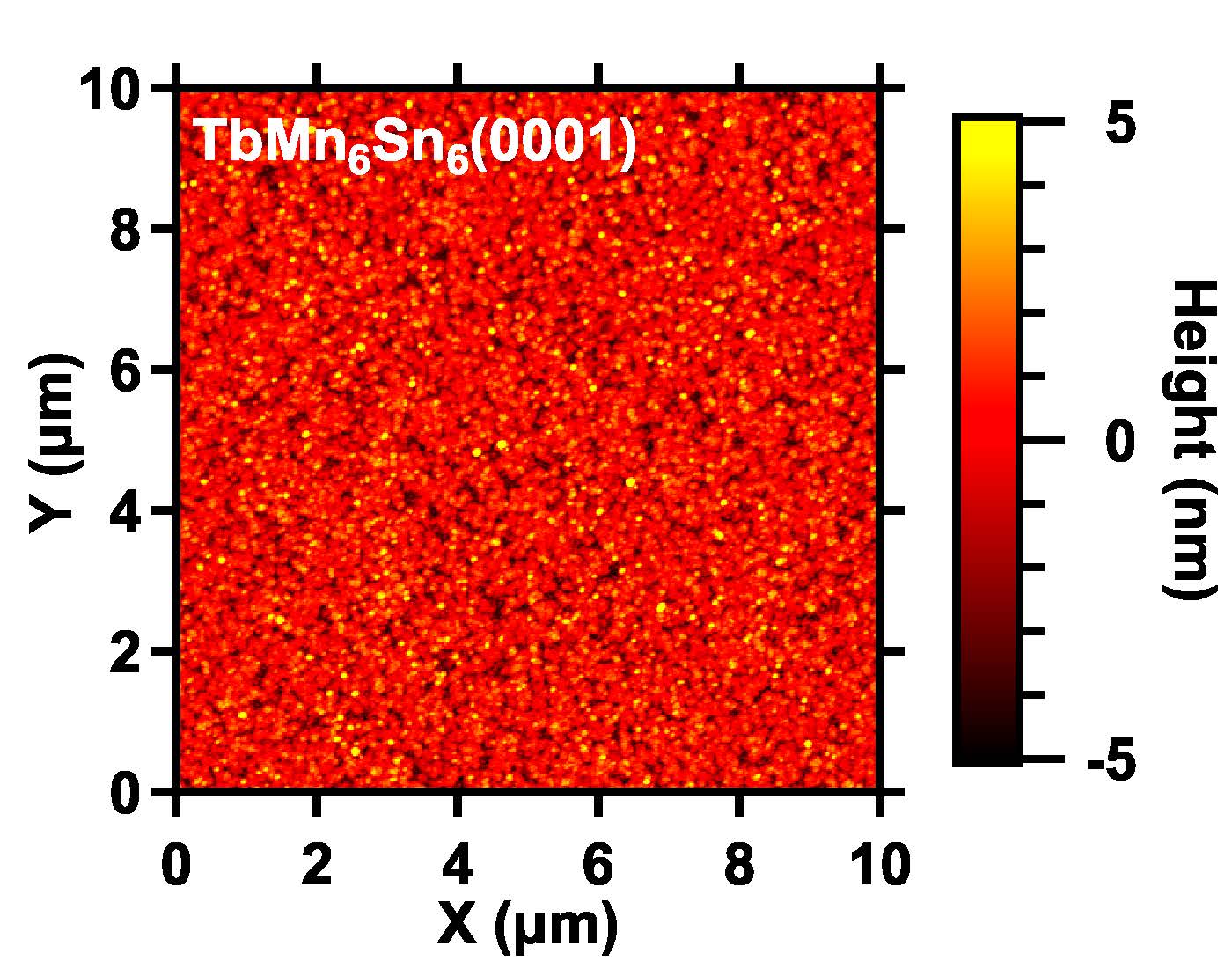}
        }
    \caption{\label{fig:Structure} Structural characterizations of RMn$_6$Sn$_6$ thin films.
    (a). XRD data of ErMn$_6$Sn$_6$(20\,nm)/Pt(5\,nm) sample.
    (b). AFM image of ErMn$_6$Sn$_6$(20\,nm)/Pt(5\,nm) sample.
    (c). XRD data of TbMn$_6$Sn$_6$(25\,nm)/Pt(5\,nm) sample.
    (d). AFM image of TbMn$_6$Sn$_6$(25\,nm)/Pt(5\,nm) sample.} 
\end{figure*}

We perform XRD measurements to confirm the crystallographic structures of ErMn$_6$Sn$_6$ and TbMn$_6$Sn$_6$ thin films.
The X-ray is from the Cu-K$\alpha$ radiation, with a wavelength of 1.5406\,\AA.
Fig.~\ref{fig:XRD_Er166} shows the XRD data of a 20\,nm ErMn$_6$Sn$_6$ thin film.
Next to the substrate peaks at 41.57$^{\circ}$ and 90.60$^{\circ}$, two major peaks are found at 39.78$^{\circ}$ and 85.46$^{\circ}$, respectively.
These peaks are attributed to Pt(111) and Pt(222) peaks because their 2$\theta$ angles convert to out-of-plane lattice constants of 2.275\,\AA~and 2.270\,\AA, respectively.
These values are close to the out-of-plane lattice constant of Pt along (111) direction ($\sqrt{3}/3\times$3.923\,\AA = 2.265\,\AA)~\cite{rooksby1964relations}.
ErMn$_6$Sn$_6$(0004) and ErMn$_6$Sn$_6$(0008) peaks appear as nearby shoulder peaks at 40.36$^{\circ}$ and 87.63$^{\circ}$, respectively.
The out-of-plane lattice constant of ErMn$_6$Sn$_6$ extracted from (0004) peak is 8.93\,\AA, which is close to but slightly smaller than the reported bulk value of 9.0006\,\AA~\cite{clatterbuck1999magneto}.

For TbMn$_6$Sn$_6$ thin film, the general characteristics of the XRD data are similar to ErMn$_6$Sn$_6$.
To the right of Pt(111) peak which is at 39.88$^{\circ}$, the TbMn$_6$Sn$_6$(0004) peak is found as a shoulder peak (see Fig.~\ref{fig:XRD_Tb166}).
A double Gaussian peak fitting yields TbMn$_6$Sn$_6$(0004) peak position of 40.26$^{\circ}$, which corresponds to an out-of-plane lattice of c = 8.95\,\AA.
This is also smaller than the reported bulk value of 9.0208\,\AA~\cite{clatterbuck1999magneto}.

We characterize the surface morphology of ErMn$_6$Sn$_6$ and TbMn$_6$Sn$_6$ thin films using AFM.
Fig.~\ref{fig:AFM_Er166} shows the AFM image of a 20\,nm ErMn$_6$Sn$_6$ sample grown at 100$^{\circ}$C.
The film has a continuous and flat surface with sub-micron grain size.
The root-mean-square (rms) roughness of the 20\,nm ErMn$_6$Sn$_6$ sample is 1.02\,nm.
Fig.~\ref{fig:AFM_Er166} shows the AFM image of a 20\,nm TbMn$_6$Sn$_6$ sample grown at 80$^{\circ}$C.
The rms roughness of the 20\,nm TbMn$_6$Sn$_6$ sample is 1.45\,nm.

\begin{figure*}[ht]
    \subfloat[\label{fig:SQUID_Er166_300K}]{
        \includegraphics[width=0.33\textwidth]{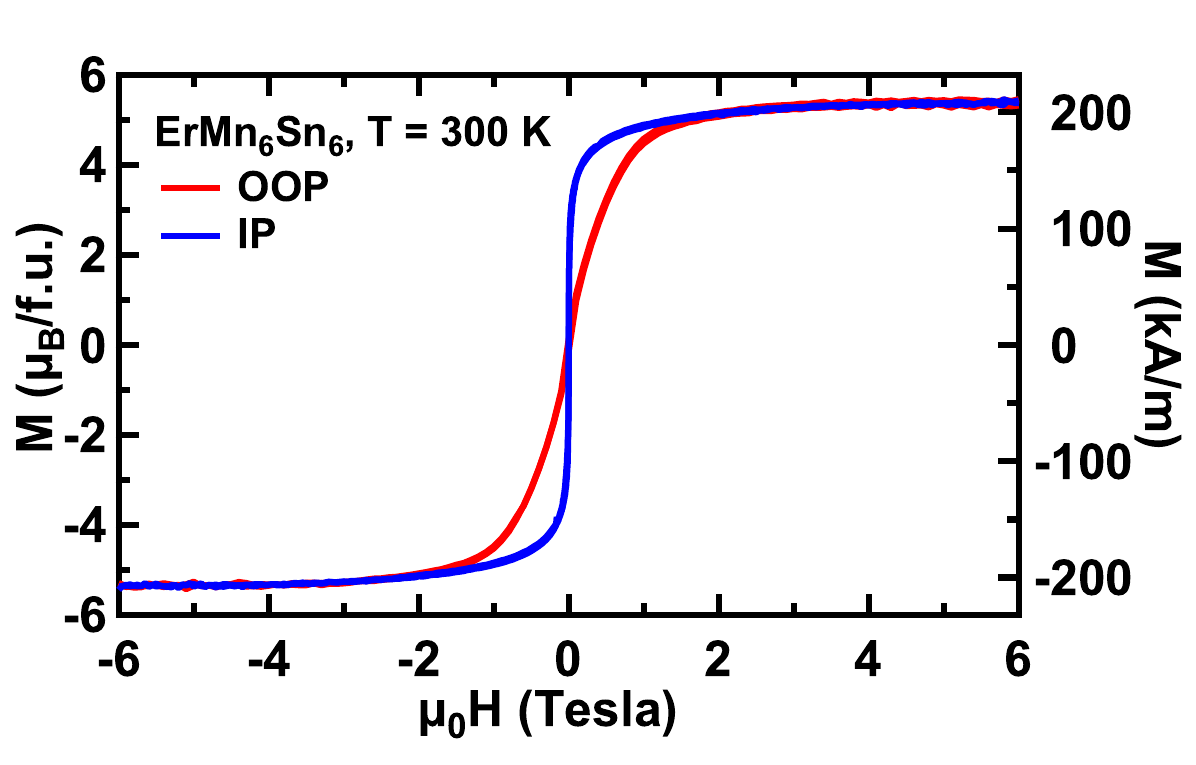}
        }
    \subfloat[\label{fig:SQUID_Er166_5K}]{
        \includegraphics[width=0.33\textwidth]{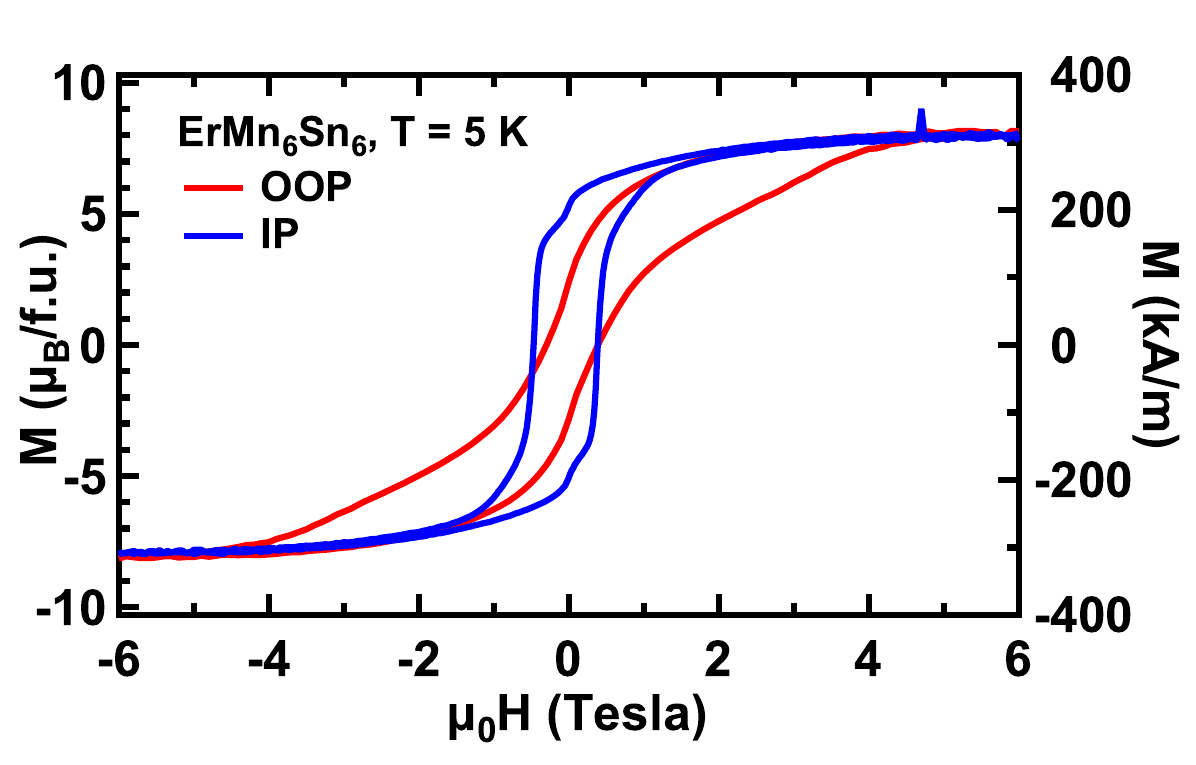}
        }
    \subfloat[\label{fig:SQUID_Er166_MvsT}]{
        \includegraphics[width=0.33\textwidth]{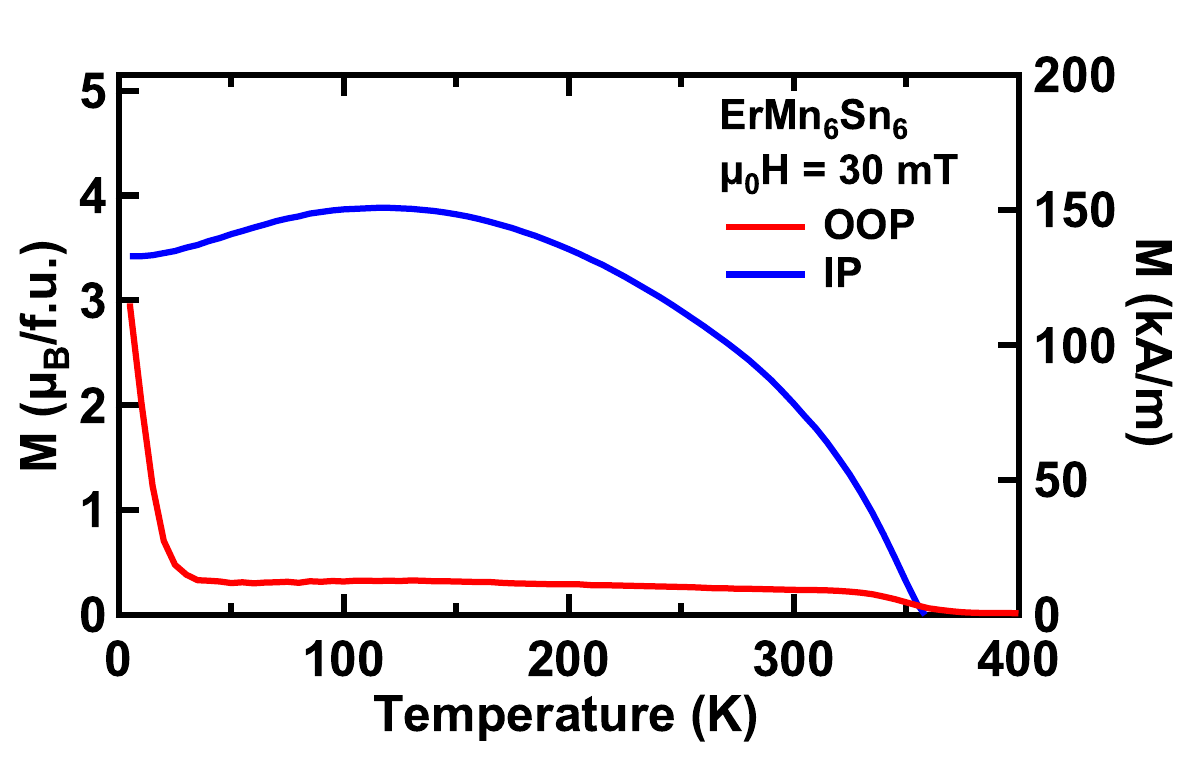}
        }\hfill
    \subfloat[\label{fig:SQUID_Tb166_300K}]{
        \includegraphics[width=0.33\textwidth]{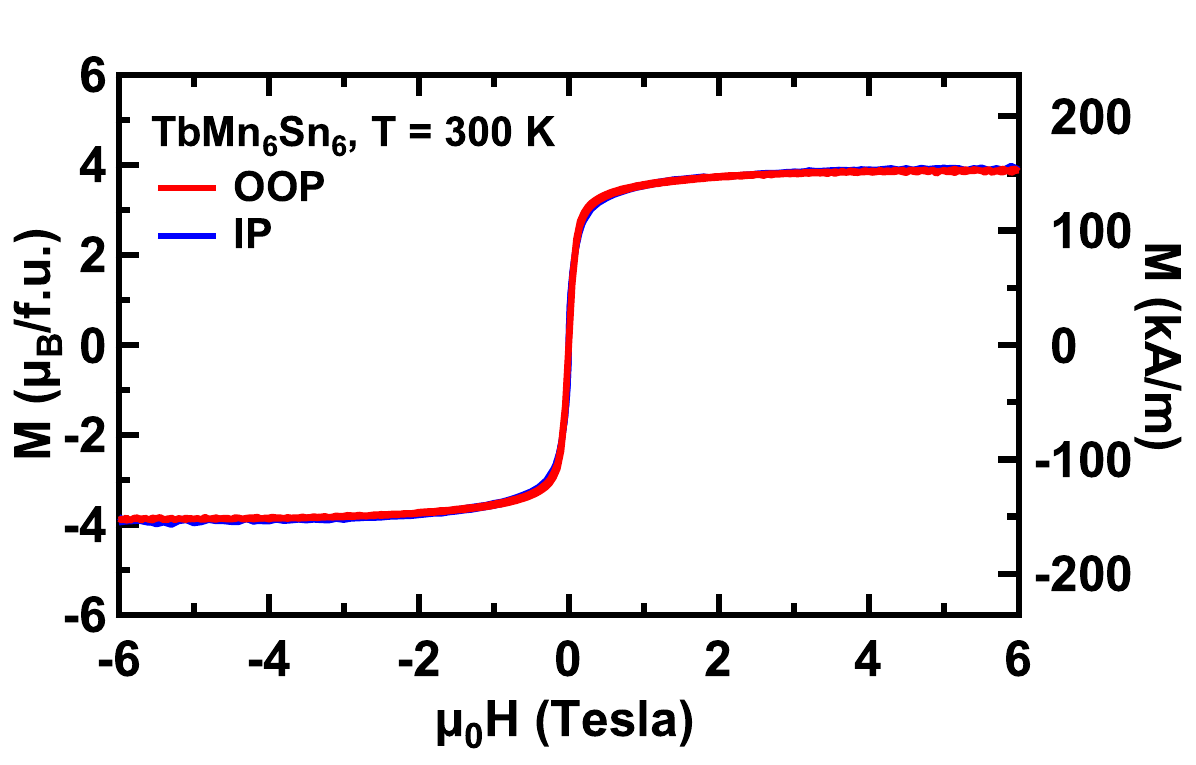}
        }
    \subfloat[\label{fig:SQUID_Tb166_5K}]{
        \includegraphics[width=0.33\textwidth]{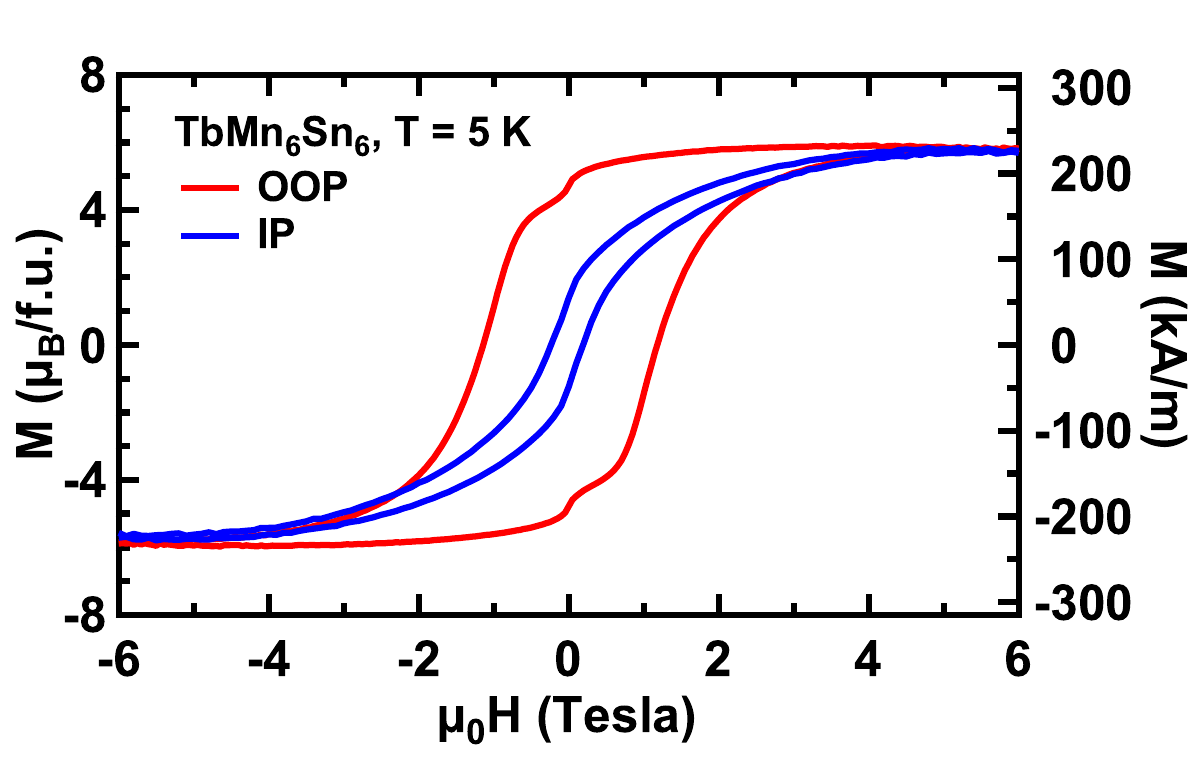}
        }
    \subfloat[\label{fig:SQUID_Tb166_MvsT}]{
        \includegraphics[width=0.33\textwidth]{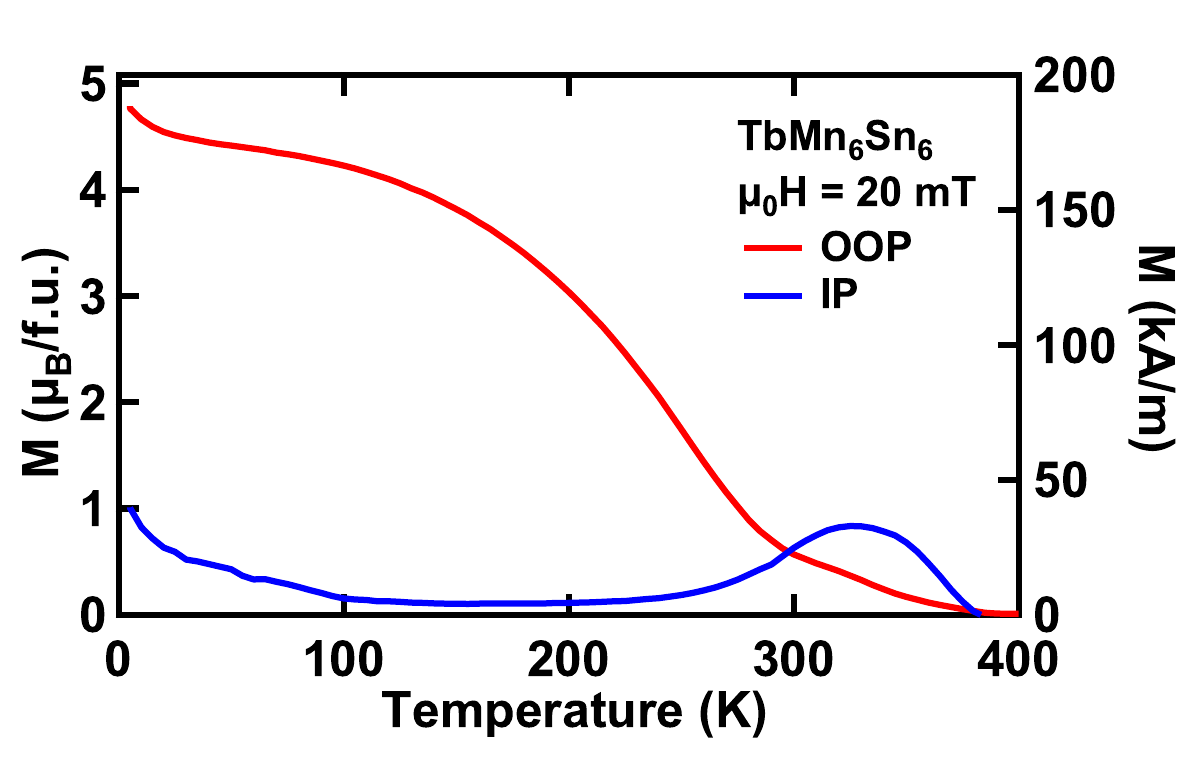}
        }
    \caption{\label{fig:SQUID} Magnetic properties of ErMn$_6$Sn$_6$ and TbMn$_6$Sn$_6$ thin films.
    (a - c). Magnetic properties of 20\,nm ErMn$_6$Sn$_6$ thin film measured with out-of-plane (OOP, red) and in-plane (IP, blue) geometries.
    (a). M vs. H hysteresis loops measured at T=300\,K.
    (b). M vs. H hysteresis loop measured at T=5\,K.
    (c). M vs. T curve.
    (d - f). Magnetic properties of 20\,nm TbMn$_6$Sn$_6$ thin film measured with out-of-plane (OOP, red) and in-plane (IP, blue) geometries.
    (d). M vs. H hysteresis loops measured at T=300\,K.
    (e). M vs. H hysteresis loop measured at T=5\,K.
    (f). M vs. T curve, showing uniaxial anisotropy for T<300\, and an OOP-to-IP spin reorientation at T=300\,K. 
    } 
\end{figure*}

To investigate the magnetic properties of the RMn$_6$Sn$_6$ thin films, we perform SQUID magnetometer measurements.
Fig.~\ref{fig:SQUID_Er166_300K} shows the IP (blue) and OOP (red) hysteresis loops of a 20\,nm ErMn$_6$Sn$_6$ thin film.
The IP hysteresis loop has a smaller saturation field of 0.2\,Tesla, while the OOP hysteresis loop has a larger saturation field of 1.7\,Tesla, indicating ErMn$_6$Sn$_6$ has easy-plane anisotropy at room temperature.
The saturation magnetization of ErMn$_6$Sn$_6$ thin film is 5.4\,$\mu_B$/f.u. (212\,kA/m) at room temperature.
The IP and OOP hysteresis loops of ErMn$_6$Sn$_6$ at T = 5\,K are shown in Fig.~\ref{fig:SQUID_Er166_5K}.
At T = 5\,K, both IP and OOP hysteresis loops show much larger coercive fields and larger saturation fields, while the saturation field of the OOP hysteresis loop is still larger than the IP hysteresis loop.
The IP loop exhibits a sharper switching behavior, with the coercive field of 0.47\,Tesla.
Meanwhile, the OOP hysteresis loop becomes narrower near zero field and wider at higher fields.
The coercive field of the OOP hysteresis loop is 0.30\,Tesla.

We also measure the temperature-dependence of remanence magnetization of the ErMn$_6$Sn$_6$ thin film, as shown in Fig.~\ref{fig:SQUID_Er166_MvsT}.
For both OOP (the red curve) and IP (the blue curve) measurements, the sample is cooled down under 7\,Tesla applied field, and measured at low field while heating up.
A 30\,mT field is applied to set the polarity of the magnetization.
From the temperature-dependent magnetization curve, the Curie temperature of the 20\,nm ErMn$_6$Sn$_6$ thin film is determined to be 353\,K, signified by a drop of magnetization to zero.
This Curie temperature is consistent with the reported value of 350\,K on ErMn$_6$Sn$_6$ bulk crystals~\cite{clatterbuck1999magnetic}.

Compared to ErMn$_6$Sn$_6$ thin films, TbMn$_6$Sn$_6$ thin films show different magnetic properties.
Fig.~\ref{fig:SQUID_Tb166_300K} and~\ref{fig:SQUID_Tb166_5K} show the hysteresis loops of a 25\,nm TbMn$_6$Sn$_6$ thin film at 300\,K and 5\,K, respectively.
At T = 300\,K, the OOP (red) and IP (blue) hysteresis loops of TbMn$_6$Sn$_6$ are almost identical, with a saturation field of 0.6\,Tesla (see Fig.~\ref{fig:SQUID_Tb166_300K}).
The saturation magnetization of TbMn$_6$Sn$_6$ is 3.9\,$\mu_B$/f.u. (154\,kA/m) at room temperature.
While cooling down to T = 5\,K, the OOP hysteresis loop of TbMn$_6$Sn$_6$ exhibits a square shape, with tails extended to 5\,Tesla, as shown in Fig.~\ref{fig:SQUID_Tb166_5K}
Meanwhile, the IP hysteresis loop shows a gradual switching behavior.
At zero field, the OOP remanence magnetization is much larger than the IP remanence magnetization, suggesting that TbMn$_6$Sn$_6$ favors OOP at T = 5\,K.

The perpendicular magnetic anisotropy is highly unusual for epitaxial kagome magnet thin films, as most of the previously reported kagome magnet thin films have easy-plane anisotropy~\cite{taylor2020anomalous, hong2020large, khadka2020anomalous, cheng2021tunable, cheng2022atomic, fujiwara2023berry}.
Even for epitaxial Fe$_3$Sn$_2$ whose bulk form has MCA along the c-axis, the MCA is overcome by the magnetic shape anisotropy (MSA) which favors easy-plane anisotropy.
As a result, both MBE-grown and sputtered Fe$_3$Sn$_2$ thin films manifest easy-plane anisotropy~\cite{khadka2020anomalous, cheng2022atomic}.
However, in TbMn$_6$Sn$_6$ thin films, the MCA is strong enough to overcome the MSA and leads to an easy-axis anisotropy at low temperatures.
It is noteworthy that such large magnetocrystalline anisotropy along the c-axis opens a gap at the Dirac cone, leading to a large intrinsic AHE~\cite{yin2020quantum, ma2021rare}.

The perpendicular magnetic anisotropy of the TbMn$_6$Sn$_6$ thin film sample at low temperatures is further confirmed by the temperature-dependent SQUID measurements, as shown in Fig.~\ref{fig:SQUID_Tb166_MvsT}.
A 20\,mT field is applied to set the polarity of the magnetization.
Below 300\,K, the OOP magnetization (red curve) is always larger than the IP magnetization (blue curve).
A crossover between IP and OOP magnetization happens at T = 300\,K above which the IP magnetization becomes larger than the OOP magnetization.
This crossover suggests an OOP-to-IP SRT as the temperature increases above 300\,K.
A similar temperature-induced SRT has been observed at 310\,K in TbMn$_6$Sn$_6$ bulk crystals~\cite{clatterbuck1999magnetic} and is reported to be related to the formation of magnetic skyrmions~\cite{li2023discovery}.
Both IP and OOP magnetization eventually disappears above the Curie temperature which is determined to be 380\,K.
This Curie temperature is lower than the reported value of 423\,K on TbMn$_6$Sn$_6$ bulk crystals~\cite{venturini1991magnetic}.

\begin{figure}[ht]
    \subfloat[\label{fig:Rxx_166}]{
        \includegraphics[width=0.4\textwidth]{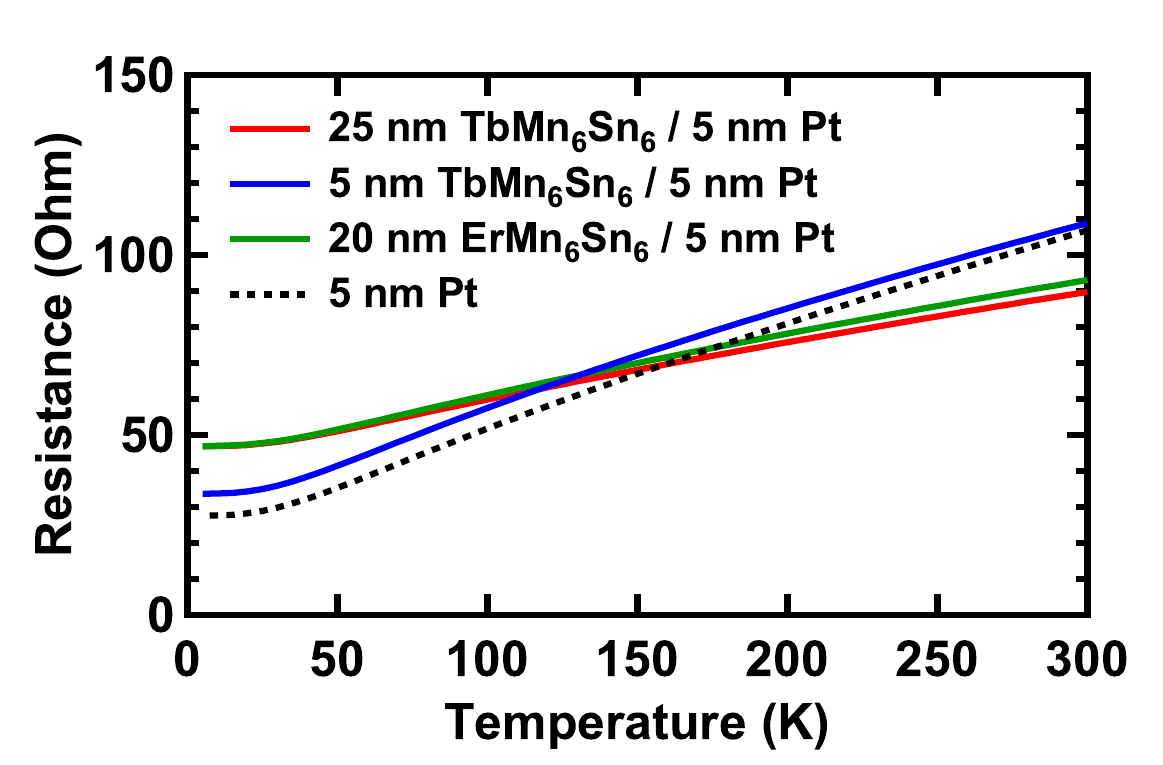}
        }\hfill
    \subfloat[\label{fig:Rxy_Er166}]{
        \includegraphics[width=0.4\textwidth]{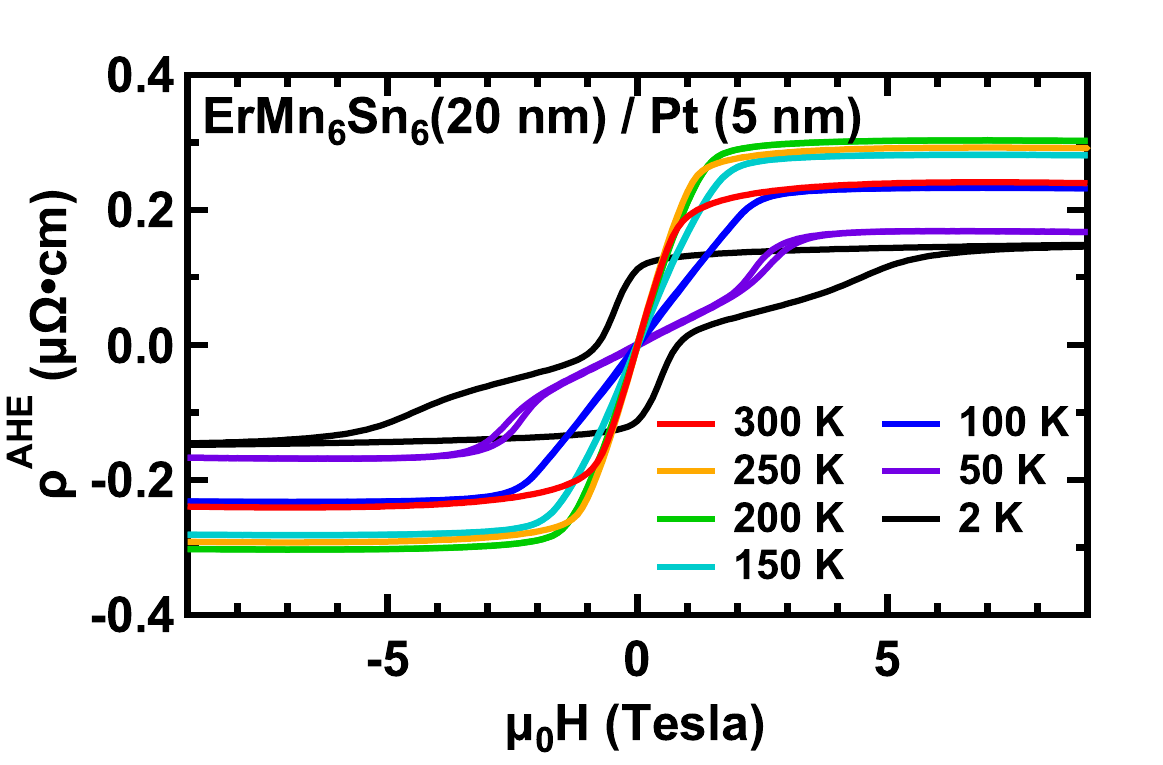}
        }\hfill
    \subfloat[\label{fig:Rxy_Tb166}]{
        \includegraphics[width=0.4\textwidth]{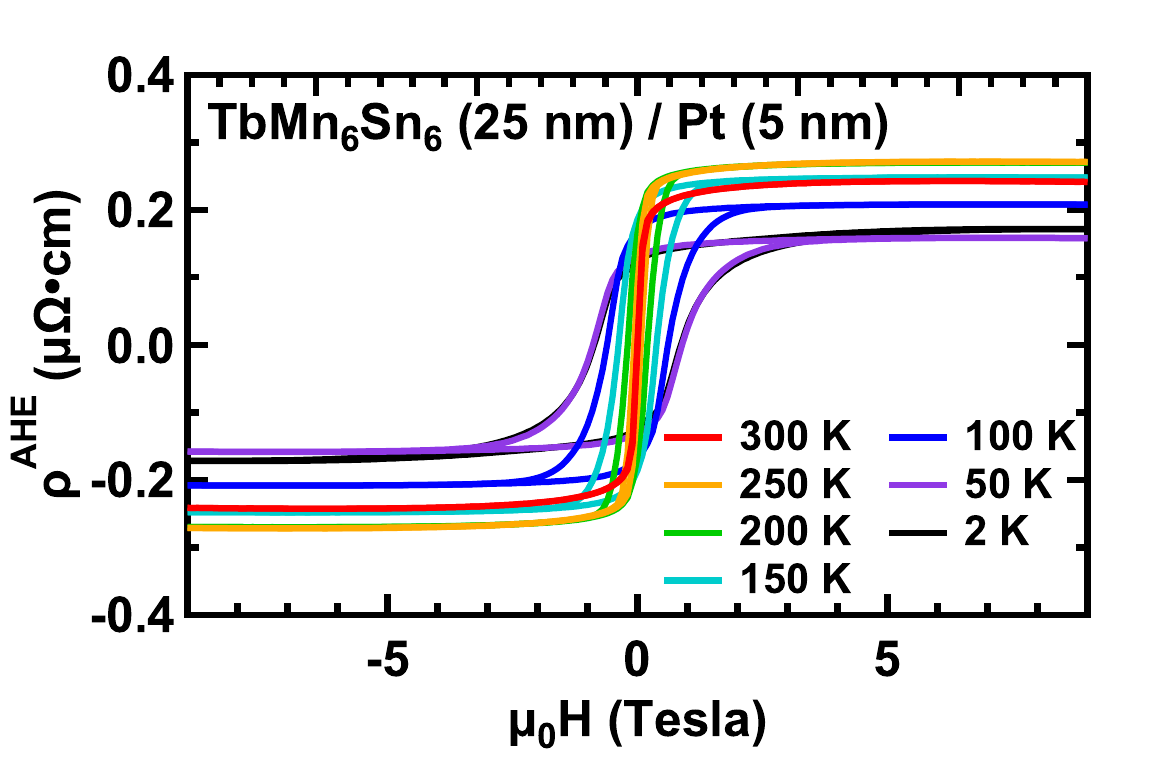}
        }\hfill
    \caption{\label{fig:Transport}  Transport measurements of ErMn$_6$Sn$_6$/Pt and TbMn$_6$Sn$_6$/Pt bilayers.
    (a). Resistance of 300\,$\mu$m$\times$100\,$\mu$m devices patterned from TbMn$_6$Sn$_6$(25\,nm)/Pt(5\,nm) bilayer (red), TbMn$_6$Sn$_6$(5\,nm)/Pt(5\,nm) bilayer (blue), ErMn$_6$Sn$_6$(20\,nm)/Pt(5\,nm) bilayer, and 5\,nm Pt single layer (black, dashed), respectively.
    (b). Anomalous Hall resistivity of ErMn$_6$Sn$_6$(20\,nm)/Pt(5\,nm) bilayer measured at different temperatures ranging from 2\,K to 300\,K.
    (c). Anomalous Hall resistivity of TbMn$_6$Sn$_6$(25\,nm)/Pt(5\,nm) bilayer measured at different temperatures ranging from 2\,K to 300\,K.
    } 
\end{figure}

We next discuss the transport properties of ErMn$_6$Sn$_6$/Pt and TbMn$_6$Sn$_6$/Pt thin films.
The transport measurements are performed using Quantum Design Physical Property Measurement System (PPMS) on 300\,$\mu$m$\times$100\,$\mu$m Hall bar devices patterned from ErMn$_6$Sn$_6$/Pt and TbMn$_6$Sn$_6$/Pt thin films.
The longitudinal resistance of the Hall bar devices at zero field is shown in Fig.~\ref{fig:Rxx_166}.
Comparing between TbMn$_6$Sn$_6$(25\,nm)/Pt(5\,nm) bilayer (the red curve) and Pt(5\,nm) single layer (the black dashed curve), the bilayer resistance is lower than the Pt single layer resistance for higher temperatures (T$>$200\,K).
This is consistent with the parallel resistor model in which the TbMn$_6$Sn$_6$ and Pt layers are modeled as two resistors connected in parallel and therefore have smaller resistance than the Pt layer alone.
The parallel resistor model has been widely used to extract the resistivity of kagome material thin films grown on metallic buffer layers~\cite{hong2020large, khadka2020anomalous, cheng2021tunable, thapaliya2021high}.
However, at low temperatures (T$<$150\,K), the resistance of the TbMn$_6$Sn$_6$(25\,nm)/Pt(5\,nm) bilayer is larger than that of the Pt(5\,nm) single layer.
This is the opposite of what one would expect from the parallel resistor model.
The comparison between ErMn$_6$Sn$_6$(20\,nm)/Pt(5\,nm) bilayer (the green curve) and Pt(5\,nm) single layer shows a similar trend that the bilayer resistance is smaller than the Pt single layer resistance for T$>$200\,K and larger than the Pt single layer resistance for T$<$100\,K.
The deviation from the parallel resistor model is more pronounced for the bilayers with smaller kagome layer thicknesses.
For example, for TbMn$_6$Sn$_6$(5\,nm)/Pt(5\,nm) bilayer (the blue curve), the resistance is larger than Pt(5\,nm) single layer from T=5\,K up to T=300\,K. 

The deviation from the parallel resistor model can be explained by Fuchs-Sondheimer theory, in which the resistance of bilayers depends not only on the bulk resistance of two materials but also the mean free path of electrons in them~\cite{fuchs1938conductivity, sondheimer1952mean}.
In the Fuchs-Sondheimer model, when the electron mean free path is much smaller than the thickness of each layer, the bilayer resistance is smaller than the resistance of individual layers.
In this case, the Fuchs-Sondheimer model converges to the parallel resistor model.
However, when the electron mean free path is much larger than the thickness of each layer, the bilayer resistance is larger than the resistance of individual layers, which is more similar to the series resistor model.
Especially in an infinite superlattice consisting of alternating stacking of two materials with the same thickness, in the thin layer limit and under the assumption that the electron densities in two materials are identical, the averaged resistivity of the superlattice is exactly equal to the average value of the bulk resistivity of the two materials~\cite{misra1999electrical}.
In our RMn$_6$Sn$_6$/Pt bilayer samples, the scattering length is shorter than the individual layer thickness at high temperatures, therefore the bilayer resistance is smaller than the resistance of Pt layer, similar to the parallel resistor model.
Meanwhile, at low temperatures or with smaller thicknesses, the mean free path becomes longer than the individual layer thickness, and the bilayer resistance is larger than the resistance of Pt layer, similar to the series resistor model.
Due to the complexity of data analysis, extracting the intrinsic longitudinal resistivity of ErMn$_6$Sn$_6$ and TbMn$_6$Sn$_6$ from bilayer resistance is outside the scope of this paper.

We also measured the anomalous Hall effect (AHE) of ErMn$_6$Sn$_6$(20\,nm)/Pt(5\,nm) and TbMn$_6$Sn$_6$(25\,nm)/Pt(5\,nm) bilayers, as shown in Fig.~\ref{fig:Rxy_Er166} and Fig.~\ref{fig:Rxy_Tb166}, respectively.
Here, the anomalous Hall resistivity is calculated by multiplying the anomalous Hall resistance by the thickness of the RMn$_6$Sn$_6$ layer.
For ErMn$_6$Sn$_6$, the AHE hysteresis loop at $T=2$\,K has a large remanence, a low-field opening (with the coercive field of $\sim 1$\,T) and a large saturation field ($\sim 6$\,T).
The opening of the hysteresis loop decreases quickly as the temperature increases, and almost disappears before the temperature reaches 100\,K.
For TbMn$_6$Sn$_6$, the AHE hysteresis loops are relatively square, with the coercive field decreasing as temperature increases.
For both materials, as temperature increases from 2\,K to 300\,K, the amplitude of AHE first increases, reaches the maximum value between 200\,K and 250\,K, and then decreases. 
This behavior is understood as follows.
The anomalous Hall resistivity $\rho^{AH}$ scales with $\rho_{xx}^{\alpha}$ where $\alpha$ is between $1$ and $2$. The intrinsic and side-jump contributions yield $\alpha=2$ and skew-scattering yields $\alpha=1$.~\cite{tian2009proper} 
Since $\rho_{xx}$ increases with temperature (Figure \ref{fig:Rxx_166}), this scaling explains the enhancement of $\rho^{AH}$ with temperature as observed below 200\,K.
On the other hand, the reduction of AHE resistivity at higher temperatures can be attributed to the decrease in magnetization as the temperature approaches the Curie temperature.

In conclusion, this study demonstrates the synthesis of $c$-plane ErMn$_6$Sn$_6$ and TbMn$_6$Sn$_6$ thin films using atomic layer molecular beam epitaxy.
A combination of RHEED, XRD, and AFM confirms the structure of the samples.
Magnetization measurements show that ErMn$_6$Sn$_6$ has easy-plane anisotropy from 5\,K to room temperature, while TbMn$_6$Sn$_6$ has perpendicular magnetic anisotropy at low temperature and an OOP-to-IP spin reorientation as the temperature increases above 300\,K.
Transport measurements at low temperatures show that the longitudinal resistance of the RMn$_6$Sn$_6$/Pt bilayer is larger than that of the Pt layer alone, which is inconsistent with the parallel resistance model. On the other hand, such behavior is qualitatively explained by the Fuchs-Sondheimer model. Lastly, since both ErMn$_6$Sn$_6$ and TbMn$_6$Sn$_6$ are synthesized under similar conditions, other members of the RMn$_6$Sn$_6$ family could likely be grown using similar AL-MBE recipes.
These results create a solid foundation for future research and applications based on RMn$_6$Sn$_6$ thin films.

\section*{Acknowledgments}

We acknowledge Daniel Halbing, Katherine Robinson, and Xueshi Gao for their assistance in the construction of the evaporation cells. 
S.C. was supported by NSF Grant No. CHE-1935885. I.L. was supported by the Center for Emergent Materials, an NSF MRSEC, under Grant No. DMR-2011876. W.Z. was supported by AFOSR MURI 2D MAGIC Grant No. FA9550-19-1-0390 and the US Department of Energy Grant No. DE-SC0016379. 

\section*{Author Declarations}

\subsection*{Conflict of Interest}

The authors declare no conflicts of interest to disclose.

\section*{Data Availability Statement}

The data that support the findings of this study are available from the corresponding author upon reasonable request.



\section*{References}

\bibliography{R166.bib}

\end{document}